\documentclass[aps,prb,floatfix,twocolumn,showpacs,preprintnumbers,amsmath,amssymb,]{revtex4-1}

\usepackage{graphicx}
\usepackage{color}
\usepackage{amsmath}
\usepackage{amssymb}
\usepackage{amsthm}
\usepackage{stmaryrd}
\usepackage{float}

\usepackage{multirow}  
\usepackage{dcolumn}   
\usepackage{bm}        
\usepackage{stmaryrd}  
\usepackage{enumerate}
\usepackage{float}

\begin{document}

\preprint{}
\title{Thermally-induced spin polarization of a two dimensional electron gas}
\author{A. Dyrda\l$^{1}$, M. Inglot$^2$, V. K.~Dugaev$^{2,3}$, and J.~Barna\'s$^{1,4}$}
\address{$^1$Faculty of Physics, Adam Mickiewicz University,
ul. Umultowska 85, 61-614 Pozna\'n, Poland \\
$^2$Department of Physics, Rzesz\'ow University of Technology,
al. Powsta\'nc\'ow Warszawy 6, 35-959 Rzesz\'ow, Poland \\
$^3$Department of Physics, Instituto Superior T\'ecnico, TU Lisbon,
av. Rovisco Pais, 1049-001, Lisbon, Portugal\\
$^4$  Institute of Molecular Physics, Polish Academy of Sciences,
ul. M. Smoluchowskiego 17, 60-179 Pozna\'n, Poland}
\date{\today }

\begin{abstract}
Spin polarization of a two-dimensional electron gas with Rashba spin-orbit interaction, induced by a thermo-current, is considered theoretically.
It is  shown that a temperature gradient gives rise to an in-plane spin polarization of the electron gas, which is normal to the temperature gradient.
The low-temperature  spin polarization changes  sign when the Fermi level crosses bottom edge of the upper electronic subband.
We also compare the results with spin polarization induced by an external electric field (current).
\end{abstract}

\maketitle


{\it Introduction} -- Spin-orbit interaction is responsible for mixing of orbital and spin degrees of freedom. This mixing, in turn, gives rise
to a variety of interesting phenomena which are observable experimentally\cite{kato04,silov04,sih05,chernyshov09,jungwirth12}. For instance, it is well known that insulating materials having sufficiently
low symmetry can reveal linear magneto-electric phenomena~\cite{fiebig05}, which follow from direct coupling of magnetic and electric degrees of freedom.
Such materials can be magnetized electrically, and the induced magnetic moment is linear in electric field. In turn, magnetic field leads then to a linear electric polarization. The situation is more complex  in  conducting materials, where electric field is inevitably  associated with a charge current.
It was predicted long time ago that the electric current in a system with spin-orbit
(SO) interaction can induce a spin polarization of conduction electrons \cite{dyakonov71},
with the polarization vector perpendicular to the direction of current and electric field.
This phenomenon was studied later in various systems exhibiting  SO interaction~\cite{edelstein90,aronov89,liu08,gorini08,wang09,schwab10}.
Edelstein~\cite{edelstein90} has considered
the spin polarization induced by electric field (current) in a two-dimensional (2D) electron gas with Rashba spin-orbit interaction, and found that the spin polarization is in the plane of the system and is normal to the electric field, as shown schematically in Fig.1 (left part). The current-induced spin polarization in a 2D electron gas for  a general case of SO interaction including both Rashba and Dresselhaus
terms has been studied theoretically  in a recent paper~\cite{wang09}.
The current-induced spin polarization has also been proved
experimentally~{\cite{kato04,silov04,sih05,yang06,stern06,koehl09,kuhlen12}.

On the other hand, it is well known that electric current can be also driven
by a temperature gradient, like for instance in the Seebeck effect. Thus,
combining the above two phenomena, one can expect that spin polarization of a system can be induced
by the gradient of temperature (by the thermo-current) as well, as shown schematically in Fig.1 (right part).
Physical origin of spin polarization, however, is now different, as the electric field and the temperature gradient affect the carrier distribution in quite different ways. As the applied electric field shifts the Fermi surface of electrons in the momentum space,
the temperature gradient makes different "smearing" of the Fermi surface at different temperatures.

\begin{figure}[h]
 \includegraphics[width=245pt]{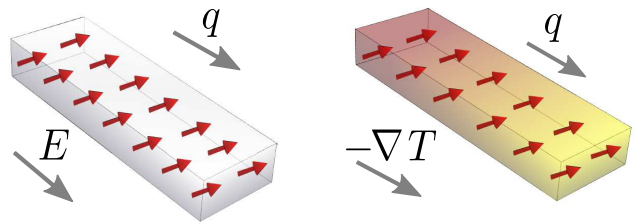}\\
 \caption{Schematic illustration of the electric-field-induced spin polarization (left side) and of
 the thermally-induced spin polarization (right side).}\label{}
\end{figure}

Spin polarization induced by a temperature gradient is in the plane of the 2D electron gas. Additionally, the spin polarization in the low temperature regime changes sign when the Fermi level crosses bottom edge of the upper subband. As the spin polarization for the Fermi level above this edge is linear in the chemical potential and spin-orbit parameter, this linearity does not hold when the Fermi level is below the edge.

{\it General considerations} -- To describe energy transport, a fictitious 'gravitational' potential has been introduced long time ago by Luttinger~\cite{luttinger64}. Gradient of this potential is a driving force for the energy current -- analogously as gradient of electrostatic potential is a driving force for charge current. It was  shown that the transport coefficients corresponding to the 'gravitational' potential coincide with those appropriate for the  temperature gradient~\cite{luttinger64}. The concept of gravitational potential was explored in a couple of papers, see e.g. Refs~\onlinecite{smrcka77,cooper97}. In this paper we use another approach. Instead of the 'gravitational' potential we introduce an auxiliary ('gravitational') vector field, which may be considered as an analog of the vector potential in electromagnetism.

Thus, we consider a system that can be described by the Hamiltonian $H = H_{0} + V$,
where $H_{0}$ includes kinetic and spin-orbit interaction terms, while $V$ is a
perturbation induced by the 'gravitational' field, which we write
in the form
\begin{equation}
V = - \hat{\mathbf{j}}_{Q}\cdot \mathbf{A}\,.
\end{equation}
Here, $\mathbf{A}$ is the 'gravitational' vector field, which  can be also understood as an auxiliary field, response to
which gives the heat current, $\hat{\bf j}_Q=-\partial H/\partial {\bf A}$.
In turn, the heat current operator $\hat{\mathbf{j}}_{Q}$ in Eq.(1) is defined as
\begin{equation}
\hat{\mathbf{j}}_{Q} = \frac{1}{2}\left[(H_{0} - \mu), \mathbf{v} \right]_{+},
\end{equation}
where $[A,B]_+=AB+BA$ and $\mathbf{v}$ is the velocity operator.

Applying the formalism to the well know results on the Seebeck effect, we may identify
$A_{i}(\omega )$ as $A_{i}(\omega ) = \frac{i}{\omega} \frac{\nabla_{i} T}{T}$.
Thus, applying the imaginary-time (Matsubara) Green function formalism,  one
can write the thermally-induced spin polarization of the system, linear in $\nabla T/T$, in the form
\begin{eqnarray}
S_{j}(\omega_m) = -\frac{1}{2} \frac{\nabla_{i} T}{T} 
\frac{iT}{\omega_m}{\rm Tr} \int \frac{d^2 \mathbf{k}}{(2 \pi)^{2}} \sum _{n}
\sigma_{j}\,
\nonumber \\ \times  G^{0}_{\mathbf{k}}(\varepsilon_n + \omega_m)
\left[(H_{0} - \mu), v_{i} \right]_{+} G^{0}_{\mathbf{k}}(\varepsilon_n),
\end{eqnarray}
where $\sigma_j$ are the Pauli matrices, $G^{0}_{\mathbf{k}}(\varepsilon_n)$ is the temperature Green's function corresponding to the Hamiltonian $H_{0}$,
the sum goes over imaginary discrete Matsubara energies $\varepsilon_n = (2n+1)i\pi T$
while $\omega_m = 2mi\pi T$ for ($m,n\in Z$). Upon calculating the sum over Matsubara energies in Eq.(3) and
making analytical continuation to the whole complex plane, one needs to take the limit $\omega\to 0$ to find the static spin polarization.\cite{abrikosov}

{\it System and solution} --
Below we apply this approach to the thermally-induced spin polarization of a 2D electron gas with Rashba SO interaction.
We will use the units with $\hbar =1$.
Hamiltonian  $H_{0}$  of the system can be written in the form
\begin{equation}
H_{0} = \varepsilon_{k} + \alpha \left(k_{y} \sigma_{x} - k_{x} \sigma_{y}\right),
\end{equation}
where $\alpha$ is the Rashba SO coupling parameter, and $\varepsilon_{k} = k^{2}/2m$.
We assume, that a temperature gradient is along the axis $y$ and calculate the spin polarization along the axis $x$ (the other components vanish). Let us consider first the case of a finite temperature, when both subbands of the electron states described by the Hamiltonian (4) are populated for arbitrary chemical potential $\mu$.
The spin polarization can be calculated from  Eq.(3), with
the Green function $G^{0}_{\mathbf{k}}$ given by
\begin{equation}
G^{0}_{\mathbf{k}}(\varepsilon_n )
= \frac{\varepsilon_n - \varepsilon_{k} + \mu + \alpha \left(k_{y} \sigma_{x} - k_{x} \sigma_{y} \right)}
{\left(\varepsilon - E_{1k} + \mu \right) \left(\varepsilon_n - E_{2k} + \mu \right) }.
\end{equation}
Here $E_{1,2k} = \varepsilon_{k} \pm \alpha k$ are the dispersion relations of the two (upper and lower) electron subbands corresponding to the Hamiltonian $H_0$.

Upon inserting Eq.(5) into Eq.(3), calculating the trace and then
calculating the sum over Matsubara energies $\varepsilon$ (by integrating over an appropriate Contour in the Complex plane), we make analytical continuation to the complex plane, and then take the limit $\omega\to 0$. As a result, one finds the dominant contribution to the static spin density $S_{x}$ in the form
\begin{eqnarray}
S_{x} =  \frac{\nabla T}{T} \int \frac{d k}{2 \pi} \tau_k \varepsilon_{k} (\varepsilon_{k} - \mu) \left[f'(E_{1k})- f'(E_{2k})\right]\nonumber\\
+ \frac{\alpha}{2} \frac{\nabla T}{T} \int \frac{d k}{2 \pi} \tau_k k (2 \varepsilon_{k} - \mu) \left[f'(E_{1k})+ f'(E_{2k})\right],
\end{eqnarray}
where $f'(\varepsilon)$ is the first derivative of the distribution function. In order to take into account relaxation processes, we have also replaced $\delta\to 1/2\tau_k$, with $\tau_k$  being the relevant relaxation time and $\delta$ an infinitesimally small number which emerges from integration over the Contour in the complex plane.\cite{abrikosov}
Since $\alpha$ is rather small, the derivative of the distribution functions can be expanded for $\alpha\ll T$, which leads to the following formula for spin density,
\begin{eqnarray}
S_{x} = 2 \alpha \frac{\nabla T}{T} \int \frac{d k}{2 \pi} \tau_k k \varepsilon_{k} (\varepsilon_{k} - \mu) f''(\varepsilon_{k}) \nonumber\\
+ \alpha \frac{\nabla T}{T} \int \frac{d k}{2 \pi} \tau_k k (2 \varepsilon_{k} - \mu) f'(\varepsilon_{k}).
\end{eqnarray}
This formula clearly shows that the leading term in spin polarization for $\alpha\ll T$ is linear in $\alpha$, and that the spin polarization
vanishes for  $\alpha =0$.

The above expansion, however,  is not valid in the limit of low temperatures, where the derivatives of the distribution functions are sharp. We derive now some approximate formula for low-$T$ spin polarization by replacing the derivatives in Eq.(6) by the corresponding delta-Dirac functions.
We will distinguish the cases of positive and negative chemical potentials, $\mu>0$ and $\mu <0$. In the former case both electron subbands are then occupied with electrons while in the later one only the lower subband, $E_{2k}$, is populated.
Let us consider first the case of $\mu>0$. From Eq.(6) one  finds the dominant contribution (linear in $\alpha$) in the form
\begin{equation}
S_{x} = \alpha\mu \frac{\nabla T}{T} \frac{m}{2\pi}\tau_{k_F}.
\end{equation}
Here, $\tau_{k_F}$ is the relaxation time at the Fermi wavevector $k_F$ corresponding to zero spin-orbit coupling, $k_F =\sqrt{2m\mu}$.
In turn, when $\mu<0$, only the lower band, $E_{2k}$, is populated, but there are two Fermi vavectors, $k_{F2}^\pm =\alpha m\pm\sqrt{\alpha^2m^2+2m\mu}$.
>From Eq.(6) one  finds then the dominant contribution in the form
\begin{equation}
S_{x} = \alpha \mu \frac{\nabla T}{T} \frac{m}{2\pi} \frac{n^{\ast}}{n}\tau_{0},
\end{equation}
where $\tau_0=\tau_{k_F=0}$ and $\mu \ge -\alpha^2m/2$. The latter condition follows from the position of the lower band edge.
Apart from this,  $n$ is the electron concentration corresponding to the Fermi level $\mu$ and $n^*$ is the corresponding electron density when
$\mu=0$.
For instance, assuming  parameters typical of GaAs-based quantum wells, i.e.,
$\alpha =2\times 10^{-9}$~eV$\cdot $cm, $k_F=10^7$~cm$^{-1}$, $\tau =10^{-11}$~s, $T=6.5$K, and $\Delta T=5$K
at the sample of length 0.1~cm, one can estimate spin polarization $S_x$ to be of the order of $S_x\simeq 2\times 10^9$~cm$^{-2}$.

{\it Numerical results} -- In the low temperature regime we have found above some approximate analytical solutions. For higher temperatures, however, we need to find the integrals in Eq.(6) or Eq.(7). To do this,  we need to know the explicit form of the relaxation time  $\tau_k $. We will consider a special case, where $\tau_k $ is constant, $\tau_k =\tau$.
Such a situation takes place for instance in the case non-ionized impurities~\cite{ando82}, where one finds
$1/\tau =4 \pi^{2} e^{4} N_{i} m/ \epsilon_{0}^{2} \kappa_{0}^{2}$. Here, $N_{i}$ is the impurity concentration, $\epsilon_{0}$ is the dielectric constant, and  $\kappa_{0}$ is the  Thomas-Fermi momentum.
The low-temperature approximate solutions are then given by  Eqs (8) and (9) with $\tau_{k_F}=\tau$ and $\tau_0=\tau$, respectively.

In the finite temperature regime, the integrals in Eqs (6) or (7) can be easily calculated by changing the integration variable from k to $\varepsilon_{k}$.
The dependence on temperature  is now more complex as additionally the distribution functions contribute to this dependence. Numerical results on the spin polarization induced by temperature gradient are shown in  Fig.2. For convenience, we normalized there the spin polarization to $\tau(\nabla T/ T)$  and the normalized spin polarization is presented as a function of the chemical potential $\mu$ for different temperatures $T$. Consider first the upper part of this figure (a), which describes the  range of positive $\mu$,  $\mu>0$, i.e. the range which is most relevant experimentally. According to Eq.(8), the spin polarization for low temperatures grows linearly with increasing $\mu$, i.e. with increasing electron concentration  (at constant $\nabla T/ T$). This linear behavior does not hold at higher temperatures (see Fig.2(a)). Apart from this, keeping constant temperature gradient one can conclude that the spin polarization decreases with increasing temperature. This is clearly shown in the inset in Fig.2(a), where to emphasize the temperature dependence, $S_x$ is normalized to $\tau\nabla T$  instead of $\tau(\nabla T/ T)$.

Part (b) of Fig.2, in turn, shows the normalized spin polarization in the range of low chemical potentials -- down to values below the bottom of the lower subband (band edge). We remind the readers, that the band edge of the upper subband $E_{2k}$ is shifted by spin-orbit interaction to a negative energy, indicated in Fig.2(b) as the band edge. The approximate solution for the low-temperature spin polarization, given by Eqs.(8) and (9), is also shown in Fig.2(b). As one can easily note, the formulas (8) and (9) describe very well the low-temperature spin polarization, except for chemical potentials very close to the band edge. Apart from this,
spin polarization at low temperatures  changes sign at $\mu=0$, and becomes negative for negative $\mu$.
When $\mu$ is below the band edge, both subbands are empty at $T=0$ (electrons are localized at the donor states), but for a nonzero $T$  some electrons are excited to the 2D subbands and a tail in the spin polarization appears for $\mu$ below the band edge (see Fig.2(b)).

\begin{figure}
\includegraphics[width=0.88\columnwidth]{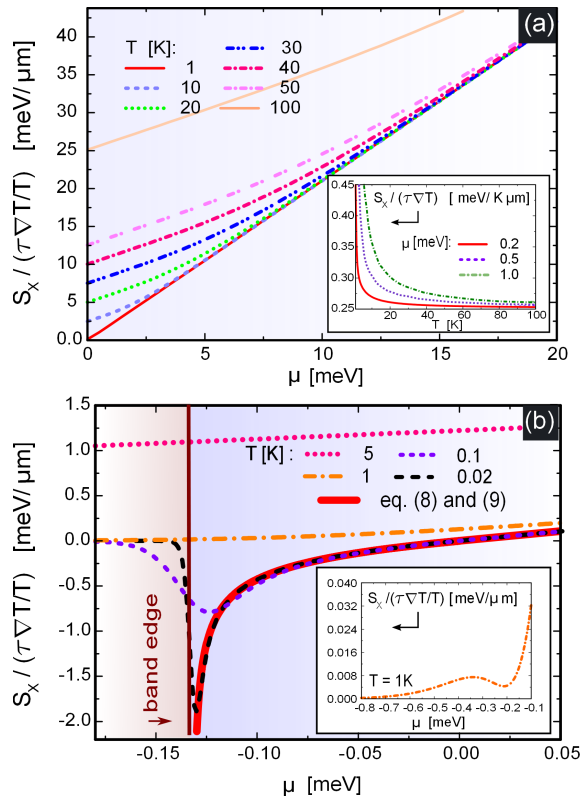}\\
  \caption{Spin polarization induced by temperature gradient, normalized to $\tau(\nabla T/ T)$ and shown as a function of chemical potential $\mu$ for indicated values of temperature. Part (a) presents the situation for $\mu>0$, while part (b) shows spin polarization for low chemical potentials, down to values below the band edge.  The inset in (a) shows the temperature dependence of the spin polarization, while the inset in (b) corresponds to $T=1$K.  The other parameters are:    $\alpha = 2\times10^{-11}$eV m, $m = 0.05m_{0}$, where $m_{0}$ is free electron mass.}\label{}
\end{figure}

The above  numerical results clearly show that spin polarization can be induced by a  temperature gradient. As mentioned already in the introductory part, the spin polarization can also be induced by a charge current (electric field), as shown already by Edelstein~\cite{edelstein90}. Below we calculate the current-induced spin polarization in the whole temperature range and for arbitrary chemical potential. This will allow us to compare the results on spin polarization induced by a temperature gradient with those obtained with an electric field. Moreover, this will also prove the theoretical method used in this paper.

{\it Spin polarization induced by electric field} --
Now we derive  some formula for spin polarization due to electric current flowing through the system. Instead of a temperature gradient, however,  there is now an electric field $\mathbf{E} = (0, E_{y}, 0)$. Accordingly, Hamiltonian (1) is now replaced with  $V = - e \mathbf{v} \cdot \mathbf{A}$, where $\mathbf{A}$ is the vector potential for electromagnetic field. Following the same methodology as above, the $x$ component of spin polarization can be written as
\begin{equation}
S_{x} (\omega_m) = - e E_{y} \frac{iT}{\omega_m}\mathrm{Tr} \int \frac{d^{2} \mathbf{k}}{(2 \pi)^{2}}
\sum _{n}
\sigma_{x} G^{0}_{\mathbf{k}}(\varepsilon_n + \omega_m) v_{y} G^{0}_{\mathbf{k}}(\varepsilon_n).
\end{equation}

\begin{figure}
 \includegraphics[width=0.88\columnwidth]{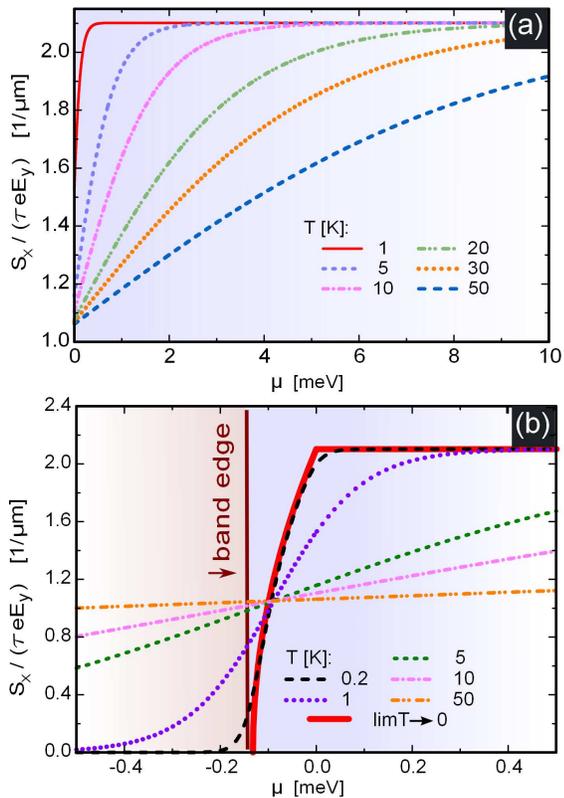}\\
  \caption{Spin polarization induced by electric field, normalized to $\tau eE_y$ and  shown as a function of chemical potential for indicated temperatures. Part (a) corresponds to positive chemical potentials $\mu >0$, while part (b) shows spin polarization for low chemical potentials, down to values below the band edge.  The other parameters are as in Fig.2.
   }\label{}
\end{figure}

For finite temperatures and constant relaxation time $\tau$, the above formula leads to the following  expression for static $S_x$:
\begin{eqnarray}
S_{x} =  eE_y\tau \int \frac{d k}{2 \pi}  \varepsilon_{k}  \left[f'(E_{1k})- f'(E_{2k})\right]\nonumber\\
+\alpha eE_y \int \frac{d k}{2 \pi}  k  \left[f'E_{1k})+ f'(E_{2k})\right],
\end{eqnarray}
which can be used for numerical calculations. In turn, the spin polarization at $T=0$,
can be obtained in a similar way as the analytical formulas (8) and (9) in the case of a temperature gradient. For
positive chemical potentials one arrives then at the following formula:
\begin{equation}
S_{x} = \alpha e E_{y} \frac{m}{2 \pi} \tau,
\end{equation}
which coincides with the  known expression for spin polarization induced by electric current \cite{edelstein90, gorini08, schwab10}.
In turn, when $\mu<0$ the corresponding formula reads
\begin{equation}
S_{x} = \alpha e E_{y} \frac{m}{2 \pi}\frac{n}{n^*} \tau .
\end{equation}

Numerical results for current-induced spin polarization are shown in Fig.3, where the  part (a) corresponds to positive $\mu$, while part (b) presents spin polarization for low chemical potentials, down to values below the band edge when the subbands are populated for nonzero temperatures.  As before, the spin polarization increases with increasing $\mu$. There is however no sign change of the spin polarization, contrary to the case of thermally induced spin polarization. Furthermore, spin polarization decreases with increasing  temperature, except for the chemical potentials in the vicinity of the band edge.

{\it Summary} -- We have calculated spin polarization of a 2D electron gas with Rashba spin-orbit coupling, induced by a temperature gradient. We have shown that the thermo-current can effectively induce spin polarization in the plane of the electron gas and normal to the direction of the temperature gradient. The method we applied is based on the concept of 'gravitational' potential, but we used the Green function formalism requiring rather an auxiliary 'gravitational' vector potential.
We have shown, that the thermally induced spin polarization changes sign when the Fermi level crosses the band edge of the upper subband.
The results also show that the thermally induced spin polarization  decreases with increasing temperature.
Additionally,  we have calculated the electric-field-induced spin polarization in the whole range of chemical potentials and for arbitrary temperature.
Contrary to the case of thermally-induced spin polarization, there is now no change of the spin polarization sign. We note that the phenomenon  studied in this paper  is different from the effects of thermally-induced spin voltage (spin thermopower or spin Seebeck effect)\cite{uchida,nunner,borge}.

This work was supported by the National Science Center in Poland as Projects Nos. DEC-2012/04/A/ST3/
00372 and partly as the Grant N N202 236540. A.D. also acknowledges support by the
Adam Mickiewicz University Foundation and by the European Union under European Social
Fund Operational Programme ''Human Capital'' (POKL.04.01.01-00-133/09-00). M.I. acknowledges support through the
project DEC-2011/01/N/ST3/00394. Two of us (A.D. and V.K.D.) acknowledge discussion with J. Berakdar and his kind hospitality.


\end{document}